\documentstyle[here,epsfig]{mn}
\begin{document}
\def\simlt{\mathrel{\rlap{\lower 3pt\hbox{$\sim$}}\raise 2.0pt\hbox{$<$}}}
\def\simgt{\mathrel{\rlap{\lower 3pt\hbox{$\sim$}} \raise
2.0pt\hbox{$>$}}}

\def\whs{\thinspace ${\rm W \: Hz^{-1} \: sr^{-1}}$}
\def\lpow{\thinspace ${\rm log_{10}P_{1.4GHz}\: }$}

\title[] 
{On the radio properties of the highest redshift quasars}
\author[M. Cirasuolo et al.] 
{M. Cirasuolo$^1$, M. Magliocchetti$^{2,4}$, G. Gentile$^{2,3}$, A. Celotti$^2$, S.
Cristiani$^4$, L. Danese$^2$  \\ 
$^1$ Institute for Astronomy, University of Edinburgh, Royal Observatory,
Edinburgh EH9 3HJ\\
$^2$ S.I.S.S.A., Via Beirut 2-4, 34014, Trieste, Italy \\
$^3$ University of New Mexico, Dept. of Physics and Astronomy, 800 Yale
Blvd Ne, Albuquerque, NM 87131, USA \\
$^4$ INAF - Osservatorio Astronomico di Trieste, Via G.B. Tiepolo 11, 40131,
Trieste, Italy
}
\maketitle 
\vspace {7cm }
\begin{abstract}
We present deep radio observations of the most distant complete quasar
sample drawn from the Sloan Digital Sky Survey. Combining our new data
with those from literature we obtain a sample which is $\sim 100$ per
cent complete down to $\rm S_{1.4GHz} = 60\mu Jy$ over the redshift
range $3.8 \le z \le 5$.  The fraction of radio detections is
relatively high ($\sim 43$ per cent), similar to what observed locally
in bright optical surveys. Even though the combined radio and optical
properties of quasars remain overall unchanged from $z \sim 5$ to the
local Universe, there is some evidence for a slight over-abundance of
radio-loud sources at the highest redshifts when compared with the
lower-z regime.

Exploiting the deep radio VLA observations we present the first
attempt to directly derive the radio luminosity function of bright
quasars at $z \simgt 4$. The unique depth -- both in radio and
optical -- allows us to thoroughly explore the population of
optically bright FR~II quasars up to $z \sim 5$ and opens a window on
the behaviour of the brightest FR~I sources. A close investigation of
the space density of radio loud quasars also suggests a
differential evolution, with the more luminous sources showing a less
pronounced cut-off at high z when compared with the less luminous ones.

\end{abstract}
%
\section{INTRODUCTION}

During the past years, studies of the properties of Active
Galactic Nuclei (AGNs) and in particular of their cosmological
evolution have become of major relevance within the more general
field of galaxy evolution. In fact, it has been found that the
properties of the central black hole (BH) are tightly related to those
of the host galaxy (e.g. Magorrian et al. 1998; Ferrarese \& Merritt
2000; Tremaine et al. 2002; McLure \& Dunlop 2004), so that the
energetic feedback that AGN activity can release is now a fundamental
ingredient in many theoretical models of galaxy formation (Silk \&
Rees 1998; Fabian 1999; Granato et al. 2001, 2004; Cavaliere \&
Vittorini 2002; Di Matteo et al. 2005; Cirasuolo et al. 2005b).

Previous studies in the optical and X-rays -- bands which are thought to 
trace the accretion processes onto the central BH -- have
established quasar and powerful AGN activity to peak at $z \sim 2$,
with a rapid decline at lower redshifts (e.g. Boyle et al. 2000; Ueda
et al. 2003; Croom et al. 2004); on the other hand, the less powerful
sources have their major shining phase at $z \simlt 1$ (e.g.
Ueda et al. 2003; Hasinger et al. 2005). The behaviour of the AGN
evolution at higher redshifts has been very uncertain for a long time
as the relevant observations were biased by selection effects and only
considered very small numbers of objects.  The advent of the
recent Sloan Digital Sky Survey (SDSS; Fan et al. 1999, York et
al. 2000) has allowed to properly explore the high redshift Universe
up to $z \sim 6$ (Fan et al. 2001a,b; 2004). These studies have 
confirmed the presence of
a cut off in the space density of quasars at $z\sim 2$.

Even though AGNs that show radio emission are only a small fraction of
the total population (Sramek \& Weedman 1980; Condon et al. 1981;
Marshall 1987; Miller, Peacock \& Mead 1990; Kellermann et al. 1989),
they represent an important subsample as the radiation at centimetre
wavelengths is unaffected by dust obscuration and
reddening. Therefore, studies of the evolution of radio-active AGNs
provide a less biased view of the behaviour of massive BHs and
accretion processes onto them as a function of cosmic time.

Several studies have addressed the evolutionary trend of radio loud
sources from the local Universe up to high z (Dunlop \& Peacock 1990;
Toffolatti et al. 1998; Jackson \& Wall 1999; De Zotti et al. 2005).
Shaver et al. (1996, 1999) argued for a drop in the space density of flat
spectrum radio quasars by more than a factor 10 between $z\sim 2.5$
and $z\sim 6$.  However, a re-analysis of such sources (Jarvis \&
Rawlings 2000)
suggests a more gradual (factor
$\sim 4$) decline, decline which is backed up by the work of Jarvis et al.
(2001) on steep spectrum sources in the same redshift interval.  
A luminosity dependent cut-off, with a decrease in space density less
dramatic for the most luminous radio sources, has been claimed by
Dunlop (1998) and confirmed by other recent studies (i.e. Vigotti et
al. 2003; Cirasuolo et al. 2005a).

Unfortunately, the process(es) responsible for the formation of radio
jets that mark the class of radio loud objects are still poorly
understood.  The mass of the central BH could play an important role
in shaping the transition between the population of radio loud
(RL) and radio quiet (RQ) AGNs. 
As recently pointed out by many
studies, RL sources seem to have the BHs confined to the upper end of the BH
mass function, whereas the BHs in RQ quasars appears to span the full range in BH
mass (Laor 2000; McLure \& Dunlop
2002; Dunlop et al. 2003; Marziani et al. 2003; McLure \& Jarvis 2004;
Metcalf \& Magliocchetti 2006). Furthermore, the analysis of a
large sample of local low luminosity AGNs drawn from SDSS showed the
fraction of galaxies hosting a radio-loud AGN to be a strong function
of BH and stellar mass (Best et al. 2005).  However, the point is
still controversial, and some authors claim no evidence for any
relation between radio power and mass of the central BH (Oshlack et
al. 2002; Ho 2002; Woo \& Urry 2002a,b; but see the dissenting view of Jarvis \&
Mclure 2002 and McLure \& Jarvis 2004, who ascribe the lack of correlation
reported by these latter authors as due to selection effects such as Doppler
Beaming and orientation).

In the light of the above discussion, the present work is aimed at
exploring the radio properties of the highest redshift quasars. The
main goal is to investigate if the physical conditions in the early
stages of galaxy formation can favour or prevent the formation of
relativistic jets and also to test if the radio loudness in quasars
exhibits some dependence on cosmic epoch.  For this purpose,
we performed deep radio observations of a sample of high redshift
quasars selected from SDSS. The optical sample is presented in Section
2, while radio observations and the radio properties of the sample are
respectively described in Sections 3 and 4. By  exploiting this unique
sample we derive the radio luminosity function in Section 5 and
investigate the behaviour of the space density of high redshift QSOs
as a function of redshift in Section 6. Our discussion and conclusions
are presented in Section 7.  Throughout this paper we adopt the
``concordance'' cosmology, consistent with the Wilkinson Microwave
Anisotropy Probe data (Bennett et al. 2003), i.e.: $\Omega_{\rm
M}=0.3$, $\Omega_\Lambda=0.7$ and $H_0=70~{\rm km~s^{-1}}$.
%
\section{The sample}
The optical quasar sample we consider in this work was drawn from the
Sloan Digital Sky Survey (SDSS) and  
represents the largest, complete, high-redshift quasar
sample to date. 

This dataset (Fan et al. 2001a) consists of 39 bright ($i^* \simlt
20$), high-redshift ($3.6 \le z \le 5$) quasars observed in the Fall
Equatorial Stripe at high Galactic latitude, covering an area of 182
sq. deg.  High redshift quasar candidates were selected using colour
cuts by exploiting the five photometric bands ($u^*$, $g^*$, $r^*$,
$i^*$, $z^*$) available from SDSS observations.

The quasar continuum is assumed to be a power law with a slope
$\alpha_{\rm o}$, i.e. $f_{\nu} \propto \nu^{-\alpha_{\rm o}}$.
The continuum magnitude, ${\rm AB}_{\rm 1450}$, is defined as the
rest-frame AB magnitude at $\lambda = 1450$ \AA\ $\:$ corrected for
Galactic extinction.  For a power law continuum, the ${\rm AB}_{\rm
1450}$ magnitude can be converted into the rest-frame Kron-Cousins B
band magnitude:
\begin{equation}\label{B_AB}
{\rm B} = {\rm AB}_{\rm 1450} + 2.5 \alpha_{\rm o} \log(4400/1450) +
0.12,
\end{equation}
where 4400 \AA\ is the effective wavelength of the B band filter, and
the factor 0.12 is the zeropoint difference between the AB magnitude
system and the Vega-based system (Schmidt et al. 1995).

The mean optical spectral index for the sources in the sample is
$\alpha_{\rm o} = 0.79$ (Fan et al. 2001a), even though Pentericci et
al. (2003) -- making use of near-IR observations (J, H and K band) --
found a slightly flatter slope, $\alpha_{o} \sim 0.5$.

Due to the optical limit $i^* \simlt 20$ adopted to select these
sources, the final sample is only representative of bright quasars
with absolute magnitudes in the range $-28.5 \simlt M_B \simlt -26$.

\section{VLA Observations}
Out of the 39 quasars which form the optical sample, 5 of them have
radio counterparts in the FIRST (Faint Images of the Radio Sky at 20
cm) survey (Becker, White \& Helfand 1995) with relatively high
1.4~GHz fluxes: $1.4 \simlt S_{\rm 1.4 GHz} \: ({\rm mJy})\simlt 10$.
These sources have an offset between radio and optical positions
of less than 0.4 arcsec, except for the faintest one (J2322-0052) which
has an offset of 0.8 arcsec, still compatible with the positional
accuracy of 1 arcsec of the FIRST survey for fluxes $\sim 1 \: {\rm
mJy}$.  Their structure is point-like, except for J0210-0018 which is
the brightest of our objects at 1.4~GHz with $S_{\rm 1.4 GHz} = 9.75
\pm 0.1 \: {\rm mJy}$; the FIRST image of this source reveals a small
component with a flux of 2 $ {\rm mJy}$ at a distance of 10 arcsec
from the centre of the optical position, probably indicating the
presence of an extended structure (with a 
proper linear size of $\sim 65$ kpc).
\begin{figure*}
\center{{
\epsfig{figure=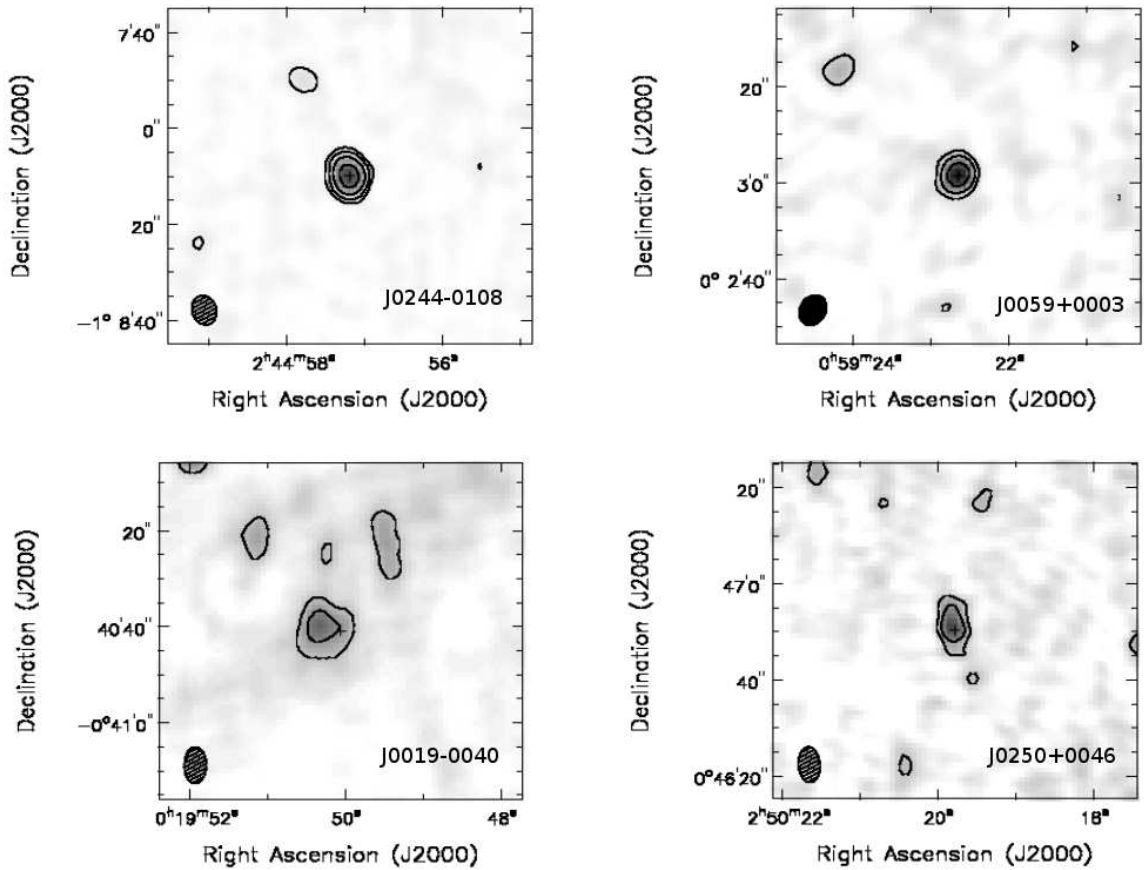,height=12cm}
}}
\caption{\label{fig_vla} Images at 1.4 GHz of the 4 high redshift
quasars detected at $> 3 \sigma_{\rm rms}$. The FWHM of the Gaussian
restoring beams are shown in the corners of all frames.}
\end{figure*}

Other 15 sources of the original Fan et al. (2001a) sample were
observed by Carilli et al. (2001), as a part of an extensive
program on high redshift quasars at radio wavelengths. These sources
were observed at 1.4 GHz for $\sim 2$ hours each, reaching a
theoretical rms noise ($\sigma_{\rm rms}$) of the order of $20-30 \: \mu$Jy.
Only 4 sources were detected at $ > 3 \: \sigma_{\rm rms}$ and other 3
sources at $2.5 \le \sigma_{\rm rms} \le 3$.  Considering the exact
coincidence of the radio positions with the optical ones, in the
following analysis we will take as true detections even these latter
three sources with $ > 2.5 \: \sigma_{\rm rms}$.

In order to constrain the radio properties of the whole optically
selected sample, we then performed deep 1.4 GHz observations with the
Very Large Array (VLA) of the remaining 19 sources.  
These observations were made in the BnA configuration in October 2003 and 
January 2004; the correlator setup was such that the total  bandwidth was 100 
MHz with two orthogonal polarizations. Every source  was observed for about 2 
hours, divided into short scans spanning a  large range in hour angle, 
in order to have a reasonably good  coverage of the u-v plane. 
The data reduction and analysis were  performed within AIPS; standard phase 
and amplitude calibration were  applied, and subsequently these were refined 
with self-calibration  using background sources in the field of view.
The dirty maps were  cleaned and then restored using Gaussian beams of about 5 
arcsec. The  theoretical noise level has been roughly achieved in most of the 
sources, except for J2306+0108 ($\sigma_{\rm rms}$ = 65 $\mu$Jy) due  to 
side-lobe confusion by a bright source in the field.

Four quasars (see Figure \ref{fig_vla}) have been detected at
$ > 3 \: \sigma_{\rm rms}$.
Two of these sources, J0244-0108 and J0059+0003, are relatively bright
($S_{\rm 1.4 GHz} = 612 \; \pm \; 20 \; \mu$Jy and $S_{\rm 1.4 GHz} =
461 \; \pm \;29 \; \mu$Jy, respectively), while the other two have
fluxes in the range $150 \simlt S_{\rm 1.4 GHz} \; (\mu {\rm Jy})
\simlt 180$ (see Table \ref{tab_vla}).  Another source (J0204-0112)
has been detected at $\sim 2.5 \: \sigma_{\rm rms}$ with a flux
$S_{\rm 1.4 GHz} = 53 \; \pm \; 20 \; \mu$Jy. The remaining 14 sources
are undetected and their fluxes will be treated as upper limits in the
following analysis.

\begin{table*}
\begin{center}
\footnotesize
\begin{tabular}{l c c c c c } \hline \hline
Source SDSS        & z  & ${\rm AB}_{1450}$ &  $M_{1450}$  & $\alpha_o$  & 
$S_{\rm 1.4 GHz} \; ({\rm \mu Jy})$ \\ \hline \hline
 J001950.06-004040.9 &  4.32 &  19.62 &   -26.36  &  -0.02 &  146    $\pm$ 39\\
 J005922.65+000301.4 &  4.16 &  19.30 &   -26.62  &  -1.09 &  461  $\pm$  29\\
 J010822.70+001147.9 &  3.71 &  19.62 &   -26.12  &  -0.19 &	$<$66	$\pm$ 33\\
 J012019.99+000735.5 &  4.08 &  19.96 &   -25.93  &  -0.52 &	$<$40	$\pm$ 20\\
 J012700.69-004559.1 &  4.06 &  18.28 &   -27.60  &  -0.66 &	$<$42	$\pm$ 21\\
 J020427.81-011239.6 &  3.91 &  19.80 &   -26.02  &  -0.83 &   53   $\pm$     20\\ 
 J020731.68+010348.9 &  3.85 &  20.10 &   -25.70  &  -1.00 &   $<$42   $\pm$ 21\\
 J023908.98-002121.5 &  3.74 &  19.60 &   -26.15  &  -0.78 &   $<$42   $\pm$ 21\\
 J024457.19-010809.9 &  3.96 &  18.46 &   -27.38  &  -1.21 &   612    $\pm$ 20\\
 J025019.78+004650.3 &  4.76 &  19.64 &   -26.49  &  -0.59 &   179    $\pm$ 19 \\
 J030707.46-001601.4 &  3.70 &  20.04 &   -25.69  &  -0.71 &   $<$36   $\pm$	 18\\
 J031036.85+005521.7 &  3.77 &  19.25 &   -26.51  &  -0.64 &   $<$36  $\pm$       18\\
 J033910.53-003009.2 &  3.74 &  19.93 &   -25.82  &  -1.17 &   $<$38   $\pm$	 19\\
 J035214.33-001941.1 &  4.18 &  19.51 &   -26.42  &  -0.16 &   $<$40   $\pm$	 20\\
 J225452.88+004822.7 &  3.69 &  20.24 &   -25.49  &  -1.51 &   $<$38   $\pm$	 19\\
 J225529.09-003433.4 &  4.08 &  20.26 &   -25.63  &  -1.15 &   $<$46   $\pm$	 23\\
 J230323.77+001615.2 &  3.68 &  20.24 &   -25.48  &  -0.77 &   $<$38   $\pm$	 19\\
 J230639.65+010855.2 &  3.64 &  19.14 &   -26.57  &  -1.38 &   $<$130  $\pm$	65\\
 J235053.55-004810.3 &  3.85 &  19.80 &   -26.00  &  -0.89 &   $<$56  $\pm$    28\\  
\hline \hline
\end{tabular}
\caption{\label{tab_vla} Properties of the 19 quasars observed at 1.4
GHz with the VLA.  Redshift and optical data are from Fan et
al. (2001a). Upper limits to the undetected sources have been defined
as $S_{\rm 1.4 GHz} = 2 \: \sigma_{\rm rms}$.}
\end{center}
\end{table*}

\section{Redshift distribution and completeness}
Combining our VLA observations with the ones made by Carilli et
al. (2001) and with the information stemming from FIRST we obtain 17 radio
detections out of the 39 quasars ($\sim 43$ per cent) constituting the
whole optical sample.

\begin{figure}
\center{{
\epsfig{figure=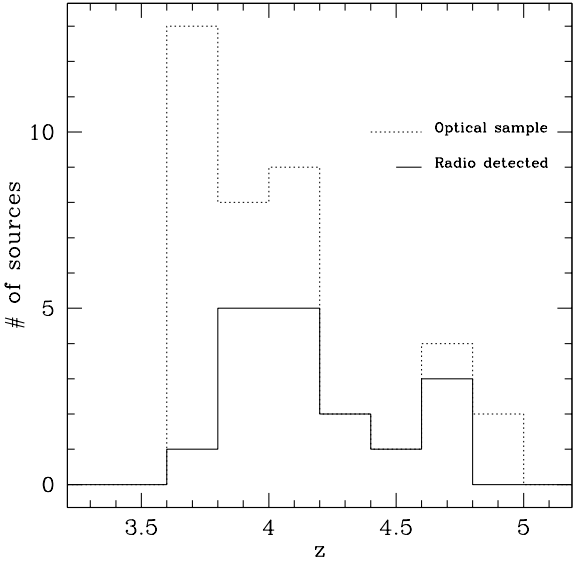,height=8cm}
\caption{\label{histoz} Redshift distribution for the radio detected
quasars (solid histogram) compared with the one for the 39 sources in
the Fan et al. (2001a) dataset (dotted histogram).}}}
\end{figure}

The redshift distribution of these radio-detected sources is shown in
Figure \ref{histoz} as a solid histogram.  For a comparison, Figure
\ref{histoz} also shows the redshift distribution of the 39 quasars of
the optical sample. It is worth noting that the fraction of radio
detections is relatively high ($\sim 40$ per cent) at $z \simgt 3.8$
but it drops to $\sim 10$ per cent in the redshift range $3.6 \simlt z
\simlt 3.8$. 
We are not aware of any specific selection effect that
can justify this rather peculiar trend.
One possible cause of
such finding would be a significant contamination of the sample by
young radio sources, whose spectrum can be inverted up to tens of GHz
and whose density is expected to increase at higher
redshifts.

Despite the results of a Kolmogorov-Smironv (KS) test, which show that the 
redshift  distribution of radio detected sources and that of the underlying 
parent  population are not statistically different ($P_{KS} \sim 0.3$), 
in order to
avoid any unforeseen bias in our analysis, in the following we will
always discuss results obtained for $z \simgt 3.8$ and only comment on
the effects of including sources from the $3.6 \simlt z \simlt
3.8$ bin.

\begin{figure}
\center{{
\epsfig{figure=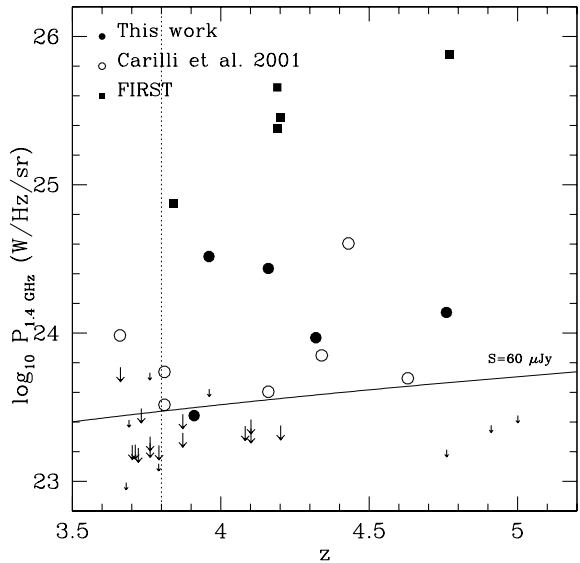,height=8cm}
\caption{\label{zp} Redshift versus radio power for the 39 sources in
the complete sample. Filled circles show detections obtained from this
work, open circles are from Carilli et al. (2001) and filled squares
are detections from FIRST. Long and short down arrows represent upper
limits from this work and from Carilli et al. (2001),
respectively. The solid line describes the selection effect due to a
radio flux limit $S_{\rm 1.4 GHz} =60 \; \mu {\rm Jy}$, while the
vertical line shows the cut at $z = 3.8$ (see text for details).}  }}
\end{figure}

As shown in Figure \ref{zp}, the radio follow up of the optical sample
in the range $3.8 \simlt z \simlt 5$ is nearly complete down to
$S_{\rm 1.4 GHz} \sim 60 \; {\rm \mu Jy}$. The only exception is a
single upper limit for the source J2309-0031 from Carilli et
al. (2001), which has a $2 \: \sigma$ detection at $S_{\rm 1.4 GHz} =
88 \pm 44 \; \mu {\rm Jy}$ with relatively high noise.  Only one
detected source (J0204-0112) lies under the flux limit of $S_{\rm 1.4
GHz} \sim 60 \; \mu {\rm Jy}$ and it is our less secure detection at a
$2.5 \: \sigma$ level. Note that pushing the limit down to $z \simlt
3.8$ would have the advantage of including another source with
$S_{\rm 1.4 GHz} > 60 \; \mu {\rm Jy}$, but would imply dealing
with three more upper limits rather than true detections.

The radio flux limit of $S_{\rm 1.4 GHz} \sim 60 \; \mu {\rm Jy}$
allows to reach completeness down to \lpow $\sim 23.6$ \whs over the
entire redshift range considered in this work. This threshold in radio
power of course depends on the adopted radio spectral
index\footnote{Throughout this work the radio flux density is defined
as ${\rm S_{\nu} \propto \nu^{-\alpha_{\rm R}}}$.},
$\alpha_{\rm R}$.  Unfortunately, to our knowledge no quasar of the
Fan et al. (2001a) sample was observed at radio frequencies
other than 1.4~GHz.  We were therefore unable to directly measure the
radio spectral index of the sources in our sample and in the following
analysis the value $\alpha_{\rm R} = 0.5$ has been adopted.
Since in the considered redshift range observations at 1.4 GHz select
the rest frame spectrum where the flat core component dominates ($\sim$ 8 GHz),
it is possible that these high-z
sources have a rather flatter radio spectral index than what assumed. 
The effects of the adoption of a flatter radio
spectral index (i.e. $\alpha_R =0$) on our results will be discussed 
throughout this paper.

\begin{figure}
\center{{
\epsfig{figure=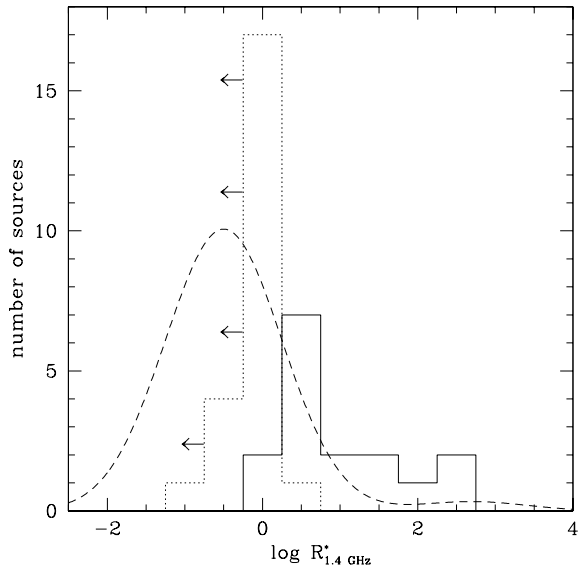,height=8cm}
\caption{\label{histor} Distribution of radio-to-optical ratios for 
radio detected sources (solid histogram) and undetected ones (dotted histogram),
for which upper limits in radio power have been used.
The dashed line is the $R^*_{1.4}$ distribution. } }}
\end{figure}

It is also interesting to look at the combined radio and optical
properties of these high z quasars. In Figure \ref{histor} the
distribution of radio-to-optical ratios\footnote{The $R^*_{1.4}$
parameter is defined as the ratio between the rest frame radio
luminosity at 1.4 GHz and the optical luminosity in the B band.}
($R^*_{1.4}$) for the radio detected sources in
the high redshift sample is shown as a solid histogram while the one derived by
using the upper limits is shown as a dotted line. For
a comparison, we also plotted as a dashed line the $R^*_{1.4}$
distribution obtained by Cirasuolo et al. (2003b) for quasars with $z
\simlt 2$, renormalized to match the number of sources in our sample.
Qualitatively, there is an overall agreement between the
$R^*_{1.4}$ distributions at high and low redshifts, at least for
$R^*_{1.4} \simlt 10$; both distributions show a region of steep transition
between the RL and RQ regime. This statement is also confirmed by
the KS test which gives high probability $P_{KS} \simgt 0.6$ that the two
distributions are not significantly different.
In the radio loud regime, instead, there is some evidence for an excess of RL
sources in the high-z dataset.  In fact, there are 5 sources with
$S_{\rm 1.4 GHz} \ge 1$ mJy in such RL tail: two of them have
$R^*_{1.4} \sim 10$, close to the RL/RQ transition region and the
other 3 have $R^*_{1.4} \simgt 100$.  By using the $R^*_{1.4GHz}$
distribution from Cirasuolo et al.  the expected number of sources
with $R^*_{1.4} \simgt 10$ (and $S_{\rm 1.4 GHz} \ge 1$ mJy) in the
same range of redshift and luminosity is $\sim 2$, suggesting an
over-abundance of RL sources at the highest z.
However, due to the limited statistics (only 5 detected objects), 
the former estimate is still
compatible with the observed number figures. In fact, a KS test applied
between the distribution of RL sources ($R^*_{1.4} \ge 10$) at $z \simgt 4$ and 
that obtained at lower
redshifts gives probabilities $0.2\simlt P_{KS} \simlt 0.5$ (depending on the
adopted value of $\alpha_R$) suggesting that the two distributions are not
statistically different. This result remains unchanged even if -- in order to
increase the statistics -- lower values of the radio-to-optical ratios 
($R^*_{1.4} \ge 2 - 3 $) are considered. 
Therefore, even if
there is some evidence for an excess of radio-loud sources at high z,
we cannot make any strong statement. Larger samples are
definitely needed to clarify this issue.

\section{Luminosity Function}\label{lf}
%
Even though small, the sample of radio detected quasars described in
the previous Section is sufficient to perform some statistical studies
and derive for the first time an estimate of the radio luminosity
function (RLF) of optically selected quasars at high
redshift. By considering both the optical and radio completeness
limits of the sample ($i^* \simlt 20$ and $S_{\rm 1.4 GHz} \ge 60 \;
{\rm \mu Jy}$, respectively), we can derive the RLF by using the
classical $1/V_{\rm max}$ method (Schmidt 1968).  For each source we
evaluated the maximum redshift at which it could have been included in
the sample, ${\rm z_{\rm max}= min(z_{\rm max}^R, z_{\rm max}^O)}$,
where ${\rm z_{\rm max}^R}$ and ${\rm z_{\rm max}^O}$ respectively correspond
to the values implied by the radio or the optical limiting flux
densities.  The incompleteness related to the optical
multicolour selection has been taken into account by applying to each
object in the sample a detection probability as provided by Fan et
al. (2001a) and used by Fan et al. (2001b) to compute the optical
luminosity function.

\begin{table}
\begin{center}
\footnotesize
\begin{tabular}{c c} \hline \hline
 \lpow  & $log_{10} \phi(P)$ \\ 
 (\whs) & $(Mpc^{-3} \Delta log_{10}P^{-1})$ \\ \hline \hline
23.9  &   $-8.03^{+0.14}_{-0.21}$ \\
       &		\\
24.6  &   $-8.12^{+0.18}_{-0.30}$ \\
       &		\\
25.3  &   $-8.40^{+0.23}_{-0.53}$ \\
       &		\\
26.0  &   $-8.87^{+0.23}_{-0.53}$ \\ \hline
\end{tabular}
\caption{\label{tab_lf} Binned radio luminosity function  
in the range $3.8 \le z \le 5$ for $\Omega_m
= 0.3$, $\Omega_{\Lambda} = 0.7$ and $H_0=70 \; {\rm km \; s^{-1} \;
Mpc^{-1}}$, as plotted in Figure \ref{rlf}. }
\end{center}
\end{table}
 
\begin{figure}
\center{{
\epsfig{figure=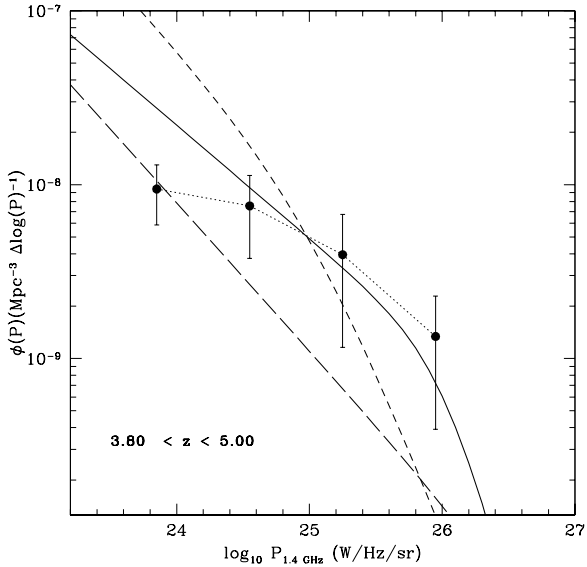,height=8cm}
\caption{\label{rlf} Radio luminosity function for sources in the
range $3.8 \le z \le 5$ with $S_{\rm 1.4 GHz} \ge 60 \; \mu {\rm Jy}$
(solid dots). For a comparison, as a solid line we show the RLF obtained by 
De Zotti et al. (2005) for flat spectrum radio quasars, calculated at the
mean $z$ of the bin.
Short dashed and long dashed curves illustrate the RLFs obtained by 
Dunlop \& Peacock (1990) for flat spectrum radio sources by respectively 
assuming  a pure luminosity evolution and a luminosity dependent evolution 
model.}  }}
\end{figure}

Figure \ref{rlf} shows the RLF obtained via the $1/V_{\rm max}$ method
for sources with $S_{\rm 1.4 GHz} \ge 60 \; \mu {\rm Jy}$ 
in the redshift range $3.8 \le z \le 5$  (see also Table \ref{tab_lf}). 
As described in the previous Section, within this redshift range the
radio follow-up is almost complete with the exception of a single
upper limit. In order to test the reliability of our estimate of the
RLF, we then re-computed it by assuming this upper limit to be a real
detection.  We find that the above assumption only affects the RLF in
a negligible way as only the faintest bin changes by a factor less
than 5 per cent. This gives us evidence for the stability of our
results.
We also computed the RLF assuming a flatter radio spectral index ($\alpha_R
=0$). The main effect is to translate the RLF towards lower radio powers by a factor
$\sim 2.5$ without affecting its shape, since the determination of the 
maximum accessible volume ($V_{\rm max}$) is driven by the optical cut. 

For a comparison, Figure \ref{rlf} also shows the high z extrapolation of the 
RLF for flat spectrum
radio quasars as recently inferred by De
Zotti et al. (2005). The
evolutionary model derived by De Zotti et al. for the population of
radio sources has been obtained by fitting number counts and redshift
distributions of several radio selected samples at different
frequencies.
An overall good agreement between our determination of the RLF and the
one derived by De Zotti et al. (2005) is found over about two
decades in radio luminosity.\\ 
In Figure \ref{rlf} we also compare our RLF estimates with the
predictions by Dunlop \& Peacock (1990) for flat spectrum radio
sources. The dashed and long dashed curves show the RLFs obtained by
respectively assuming a
pure luminosity evolution and a luminosity dependent evolution model, 
translated to the concordance cosmology adopted in this paper.
We note a moderate agreement with observations only for
\lpow $\simlt 25$ \whs in the case of luminosity dependent evolution of radio
sources. However, 
it is worth noticing that the Dunlop \& Peacock  RLFs 
are completely unconstrained at these high 
redshifts, so they  have to be merely considered as a pure extrapolation. 


It is however important to notice that a direct comparison between the
RLF obtained from our high redshift quasars and models for the
different radio populations as derived by De Zotti et al. (2005)
or Dunlop \& Peacock (1990) is not straightforward. These models in
fact include all the radio sources regardless of their optical
properties, 
while our RLF is only representative of bright ($M_B < -26.5$) optical
quasars. 
For a meaningful data-to-model comparison we should than 
take into account the (unknown)
contribution from fainter FRII sources and quasars. 
The problem is however not as bad as it looks since -- as
Cirasuolo et al. (2003b) have shown -- even though with a large
scatter, optical and radio luminosities are broadly correlated
so that the contribution of these "missing'' faint sources should
typically affect the RLF at luminosities lower than those probed by
our work.

As we will discuss in more detail in Section \ref{dens}, we have also
compared our findings with extrapolations of the RLF obtained at lower z 
by Cirasuolo et al.  (2005a) and Willott et al. (2001). These
extrapolations are not able to reproduce the high redshift data
presented in this paper since they assume a negative evolution which is too 
quick for $z \simgt 2-3$.

\section{Space Density}\label{dens}
%
The number of quasars in our sample only allowed us to compute the RLF
in a single redshift bin.
Moreover, as described in the previous Section,
comparisons with evolutionary models for the different radio
populations is biased by the bright optical cut in our sample and the
lack of information on the radio spectral indices.  Therefore, the
best way to explore the cosmological evolution of radio quasars and
shed light on the behaviour of radio loudness up to the highest
accessible redshifts is to look at the evolution of an integrated quantity such
as their space density.

\begin{figure}
\center{{
\epsfig{figure=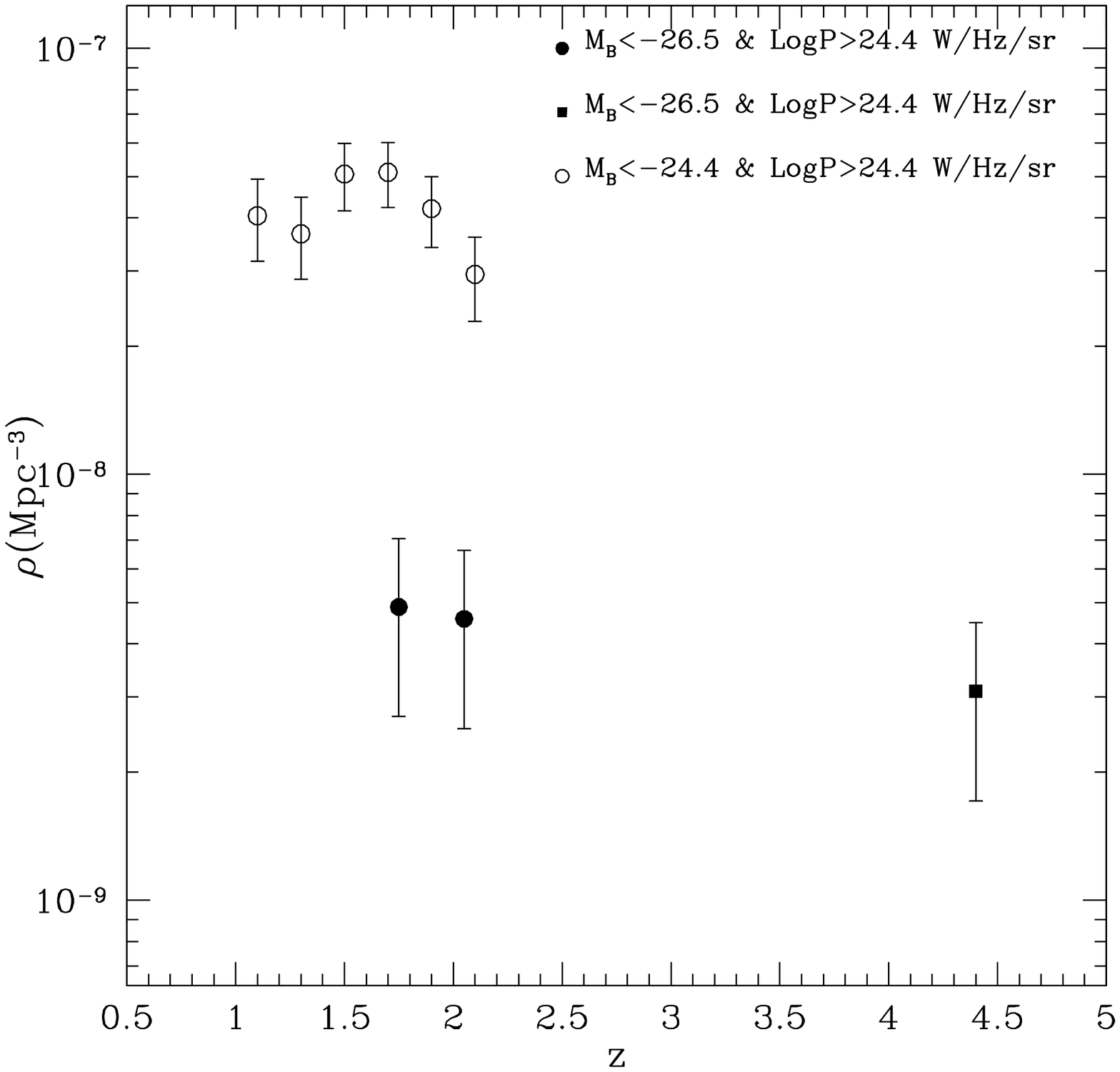, height=8cm}
\caption{\label{cumul} Space density of radio quasars with \lpow $\ge
  24.4$ \whs computed at different redshifts and for different optical
  luminosities (see text for details). The filled square represents
  the space density for those sources in the high redshift sample
  ($3.8 \le z \le 5$) presented in this paper ($M_B \le -26.5$), 
  while the filled dots show the
  space density for the combined 2dF-LBQS-FIRST sample at lower
  redshifts derived by applying the same optical and radio cuts as
  above. The space density of optically fainter sources ($M_B \le
  -24.4$) from the combined 2dF-LBQS-FIRST sample as obtained by Cirasuolo et al. 
  (2005)  is shown at the
  different redshifts as open circles.}  }}
\end{figure}

Figure \ref{cumul} shows the comparison between space densities of
radio quasars as computed at different redshifts and for different
ranges of optical luminosities.  The solid square in the Figure
represents the space density obtained from the high redshift sample
($3.8 \le z \le 5$) presented in this work by considering sources with
$M_B \le -26.5$ and \lpow $\ge 24.4$ \whs.  The optical cut is due to
the apparent magnitude limit $i^* \simlt 20$ of the optical
sample. The cut in radio power has instead been chosen in order to
have an unbiased comparison with the space densities computed at lower
redshifts.  For $z \simlt 2.2$, the space density was derived by using
a combined sample from the 2dF Quasar Redshift Survey (Croom et
al. 2004) and the Large Bright Quasar Survey (LBQS; Hewett et
al. 1995). Both these surveys have been cross-correlated with the
FIRST dataset (Cirasuolo et al. 2003a, 2005; Hewett et al. 2001) in
order to obtain samples of quasars complete in radio down to a flux
limit of $S_{\rm 1.4 GHz} = 1 \; {\rm mJy}$, which translates into
\lpow $\simgt 24.4$ \whs at $z \simlt 2.2$ (open circles).
In order to minimize selection effects in the comparison of the space
densities of high and low redshift quasars, we then applied to the low
redshift sample the same optical ($M_B \le -26.5$)
cut of the $z\simeq 4$ dataset.
It is worth noting that some selection effects can arise from the different
selection techniques applied to construct the 2dF-LBQS samples and that of 
SDSS quasars. However, at the bright magnitudes considered in this paper these 
effects should be negligible.

The filled dots in Figure \ref{cumul} show the space density obtained
from the combined 2dF-LBQS-FIRST sample as described above.  It is
clear from the Figure that the drop in the space density of optically
bright RL quasars between $z\sim 2$ and $z \sim 4.4$ is {not very pronounced}
although errors are large. This result remains
mostly unchanged when the space density is computed over the entire
redshift range $3.6 \le z \le 5$. In fact, no other source with \lpow
$\ge 24.4$ \whs $\;$ is present in the range $3.6 \le z \le 3.8$, and the
effect of sampling a slightly larger volume is minimal.
 
To quantify the decline in the space density of optically bright 
radio quasars we apply a linear regression to 
the three data
points with $M_B < -26.5$, 
obtaining a value for the slope
of $-0.07 \pm 0.11$, entirely consistent with no evolution. 
If we instead use a flatter radio spectral index ($\alpha_R \sim 0$) for the
high redshift sources,
the space density 
at $z \sim 4.4$ drops by a factor $\sim 1.5$ leading to a slightly steeper
slope of $-0.14 \pm 0.13$. Therefore, we conclude that, independent of the
adopted radio spectral index, the drop in the space density of optically
bright RL quasars between $z\sim 2$ and $z \sim 4.4$ is at most a factor
$\sim 1.5-2$.

The above trend is in agreement with recent results from Vigotti et
al. (2003).  By using a sample of 13 radio quasars, these authors
showed the decline in their space density to be approximately a factor
2 between $z \sim 2$ and 4. This is close to our findings, even though
the objects used by Vigotti et al. (2003) are about one magnitude
brighter in optical ($M_B \simlt -27.5$) than those in our sample, 
while the radio selection in the two datasets is roughly the same
(\lpow $> 24.6$ \whs).

It is interesting to note that the behaviour of bright quasars seems
to differ from that of fainter sources which instead present a
sharper decline toward higher redshifts.  Figure
\ref{cumul} shows (open circles) the space density at low
redshifts ($z \simlt 2.2$) of fainter ($M_B \le -24.4$, completeness
limit of the combined 2dF-LBQS-FIRST dataset for $z \simlt 2.2$)
optical quasars.  As already pointed out by Cirasuolo et al. (2005a),
this space density shows a decrement of a factor $\sim 2$ already in the narrow
redshift range $z
\sim 1.7$ and $z \sim 2.2$. 
In fact, by fitting with a straight line the values
of the space density at $z \ge 1.7$ we find a slope of $-0.6 \pm 0.3$.
This trend, when compared with the
behaviour of the optically brighter sources which only show a
moderate decline up to $z \sim 4.4$, is suggestive of a
differential evolution of radio-active sources of different
optical  luminosities.  Certainly, the lack of information on
the space density of fainter sources at higher redshifts ($z \simgt
2.2$) does not allow any definitive conclusion, but our results 
indicate the cosmological evolution of radio activity in quasars to
be a function of their optical power.

\section{Discussion and Conclusions}

We have presented deep radio observations of the most distant
complete quasar sample drawn from the Sloan Digital Sky
Survey. Combining our deep VLA observations with the ones performed by
Carilli et al. (2001) and also with the 5 detections from FIRST, 
we obtained $\sim
100$ per cent completeness down to $S_{\rm 1.4 GHz} = 60 \; {\rm \mu
Jy}$ over the redshift range $3.8 \le z \le 5$.

The fraction of radio detections is relatively high ($\sim 43$ per
cent), similar to what observed locally for bright optical surveys
such as the Palomar Bright Quasar Survey (Kellermann et al. 1989) and
the LBQS (Hewett et al. 2001). A
comparison between the $R^*_{1.4}$ distribution of these high redshift
radio quasars with the one derived at $z \simlt 2$ by Cirasuolo et
al. (2003b) suggests that the combined radio and optical properties of
quasars might remain overall unchanged from $z \sim 5$ to the
local Universe. However, even though the shape of the $R^*_{1.4}$
distribution is roughly preserved over cosmic time, there is some
marginal evidence for a slight over-abundance of radio loud sources
at high z when compared with the low redshift samples (see Figure
\ref{histor}), even though not statistically significant due to the small
number of sources. 
Furthermore, it is
worth noting that the adoption of a flatter $\alpha_{\rm R}$ for high z objects
would reduce their radio power and shift them
towards lower values of $R^*_{1.4}$, therefore somehow reducing the
fraction of purely RL ($R^*_{1.4}$) sources.

An interesting hint to shed some light on the above issue comes from
comparisons with the parent optical population. The space density of
bright optical quasars ($M_B < -26.5$) at $z\sim 2$ and $z \sim 4.4$
is $\rho_O \sim 2 \times 10^{-7} \rm Mpc^{-3}$ and $\rho_O \sim 1.5
\times 10^{-8} \rm Mpc^{-3}$, respectively (see Fan et al. 2001b,
2004).  The ratio between the space densities of radio sources with
\lpow $\ge 24.4$ \whs (as plotted in Figure \ref{cumul}) and the
$\rho_O$ of the total optical population is therefore $0.025 \pm 0.01$
at $z \sim 2$ and $0.15 \pm 0.1$ at $z \sim 4.4$. These figures have been
obtained for a radio spectral index $\alpha_R=0.5$ but, as shown in section 6,
the results are not expected to exhibit great variations by adopting a flatter
slope. Again, this
suggests that at high redshifts the probability of having RL sources
is enhanced with respect to that at lower redshifts.
However, we stress once
more that the statistics we dealt with in this work is very poor and 
further data and larger
samples are needed in order to have a more robust answer. \\ Even
though a detailed investigation of this phenomenon is outside the
possibility of the present data, in a very qualitative way we
can relate the suggested excess of RL sources at high $z$ as
compared with the lower redshift regime with changes in the accretion
rates with cosmic time. 
The physical conditions of the primordial massive galaxies hosting
quasars and the availability of a larger amount of gas in these early
stages could in fact allow super-Eddington accretions and eventually
favour the formation of powerful radio jets. 

We also attempted the first direct estimate of the radio
luminosity function of bright quasars at $z \simgt 4$. Exploiting the
deep radio flux limits obtained through VLA observations, we were able
to trace the RLF down to \lpow $\sim 23.6$ \whs. It is worth noticing
that the transition region between the FR~I and FR~II population
occurs at \lpow $\sim 24.4$ \whs.  Therefore, the unique depth -- both
in radio and optical -- of this high redshift quasar sample allows us
to completely explore the population of optically bright FR~II quasars
up to $z\sim 5$ and furthermore opens a window on the behaviour of the brightest
FR~I sources. 

Finally, close investigation of the RL quasar space density at
different redshifts is suggestive of a differential
evolution for the two populations of optically faint and bright
objects.  The more luminous sources in fact show a less pronounced
cut-off at high z -- with a drop in their space density of only a
factor $\sim 2$ between $z \sim 2$ and $z \sim 4.4$ -- when compared
with the less luminous ones. Even though the lack of information on
the behaviour of optically faint quasars at $z \simgt 2.2$ does not
allow any definitive conclusion, our results indicate the
cosmological evolution of radio activity in quasars to be a function
of their optical power.

\section{acknowledgements}

MC acknowledges the support of PPARC, on rolling grant no. PPA/G/O/2001/00482.
AC and LD acknowledge the Italian MIUR for financial support.


\begin{thebibliography}{} 

\bibitem{} Becker R.H., White R.L., Helfand D.J., 1995, ApJ, 450, 559
\bibitem{} Bennett C.L., et al. 2003, ApJ,  583, 1
\bibitem{} Best P.~N., Kauffmann G., Heckman T.~M., Brinchmann J., Charlot S., Ivezi{\'c} {\v Z}.,White, S.~D.~M., 2005, MNRAS, 362, 25 
\bibitem{} Boyle B.J., Shanks T., Croom S.M., Smith R.J., Miller L., Loaring N., Heymans C., 2000, MNRAS, 317, 1014
\bibitem{} Carilli C.L., Bertoldi F., Rupen M.P., Fan X., Strauss M.A., et al.,  2001, ApJ, 555, 625
\bibitem{} Cavaliere A. \& Vittorini V., 2002, ApJ, 570, 114
\bibitem{} Cirasuolo M., Magliocchetti M., Celotti A., Danese L., 2003a,  MNRAS, 341, 993
\bibitem{} Cirasuolo M., Celotti A.,  Magliocchetti M., Danese L., 2003b, MNRAS, 346, 447
\bibitem{} Cirasuolo, M., Magliocchetti, M.,  Celotti, A., 2005a, MNRAS, 357, 1267 
\bibitem{} Cirasuolo, M., Shankar F., Granato G.L., De Zotti G., Danese L., 2005b, ApJ, 629, 816
\bibitem{} Condon J.J., O'Dell S.L., Puschell J.J., \& Stein W.A. 1981, ApJ, 246, 624
\bibitem{} Croom S.M., Schade D., Boyle B.J., Shanks T.,  Miller L., Smith R.J., 2004, ApJ, 606, 126
\bibitem{} De Zotti, G., Ricci, R., Mesa, D., Silva, L., Mazzotta, P., Toffolatti, L., Gonz{\'a}lez-Nuevo, J., 2005, A\&A, 431, 893 
\bibitem{} Di Matteo T., Springel V., Hernquist L., 2005, Nature, 433, 604
\bibitem{} Dunlop J.S. \& Peacock J.A., 1990, MNRAS, 247, 19 
\bibitem{} Dunlop J.S., 1998, in Bremer M.N., et al., eds., Observational Cosmology with the New Radio Surveys. Kluwer
\bibitem{} Dunlop J.S., McLure R.J., Kukula M.J., Baum S.A., O'Dea C.P., Hughes D.H., 2003, MNRAS, 340, 1095
\bibitem{} Fabian A.C. 1999, MNRAS,  308, L39
\bibitem{} Fan X., Strauss M.A., Schneider D.P., Gunn J.E., Lupton R.H., 1999, AJ, 118, 1
\bibitem{} Fan X., Strauss M.A., Richards G.T., et al. 2001a, AJ, 121, 31
\bibitem{} Fan X., Strauss M.A., Schneider D.P., et al. 2001b, AJ, 121, 54
\bibitem{} Fan X., Hennawi J.F., Richards G.T., et al.,  2004, AJ, 128, 515	
\bibitem{} Ferrarese L. \& Merritt D., 2000, ApJ, 539, L9
\bibitem{} Granato G.L., Silva L., Monaco P., Panuzzo P., Salucci P., De Zotti G., Danese L.,  2001, MNRAS, 324, 757
\bibitem{} Granato G.L., De Zotti G., Silva L., Bressan A., Danese L.,  2004, ApJ, 600, 580
\bibitem{} Hasinger G., Miyaji T., Schmidt M., 2005, A\&A, 441, 417
\bibitem{} Hewett P.C., Foltz C.B., Craig B., Chaffee F.H., 1995, AJ, 109, 1498
\bibitem{} Hewett P.C., Foltz C.B., Chaffee F.H., 2001, AJ, 122, 518
\bibitem{} Ho L.~C., 2002, ApJ, 564, 120
\bibitem{} Jackson C.A. \& Wall J.V., 1999, MNRAS, 304, 160
\bibitem{} Jarvis M.J., Rawlings S., 2000, MNRAS, 319, 121
\bibitem{} Jarvis M.J., et al. 2001, MNRAS, 327, 907
\bibitem{} Jarvis M.J. \& McLure R.J., 2002, MNRAS, 336, 38
\bibitem{} Kellermann K.I., Sramek R., Schmidt M., Shaffer D.B., Green R, 1989, AJ, 98, 1195
\bibitem{} Laor A., 2000, ApJ, 543, L111 
\bibitem{} Magorrian J., Tremaine S., Richstone D., Bender R., Bower G., Dressler A., Faber S. M., Gebhardt K., Green R. Grillmair C., et al. 1998, AJ, 115, 2285
\bibitem{} Marshall H.L., 1987, ApJ, 316, 84
\bibitem{} Marziani P., Zamanov R.K., Sulentic J.W., Calvani C., 2003, MNRAS, 345, 1133
\bibitem{} McLure R.J. \& Dunlop J.S., 2002, MNRAS, 331,795 
\bibitem{} McLure R.J. \& Dunlop J.S., 2004, MNRAS, 352,1390 
\bibitem{} McLure R.J. \& Jarvis M.J., 2004, MNRAS, 353, L45
\bibitem{} Miller L., Peacock J.A.,  Mead A.R.G., 1990, MNRAS, 244, 207
\bibitem{} Oshlack A.~Y.~K.~N., Webster R.~L., Whiting M.~T. 2002, ApJ, 576, 81 
\bibitem{} Pentericci, L., et al., 2003, A\&A, 410, 75 
\bibitem{} Schmidt M., 1968, ApJ, 151, 393
\bibitem{} Schmidt M., van Gorkom J.H., Schneider D.P., Gunn J.E.,  1995, AJ, 109. 473
\bibitem{} Shaver P.A., Wall J.V., Kellermann K.I., Jackson C.A., Hawkins M.R.S., 1996, Nature, 384, 439
\bibitem{} Shaver P.A., Windhorst R.A., Madau P., de Bruyn A.G., 1999, A\&A, 345, 380
\bibitem{} Silk J. \& Rees M.J., 1998, A\&A, 331, L1
\bibitem{} Sramek R.A. \& Weedman D.W., 1980, ApJ, 238, 435
\bibitem{} Toffolatti L., Argueso Gomez F., De Zotti G., Mazzei P., Franceschini A., Danese L., Burigana C., 1998, MNRAS, 297, 117
\bibitem{} Tremaine S., et al. 2002, ApJ,  574, 740
\bibitem{} Ueda Y., Akiyama M., Ohta K., Miyaji T., 2003, ApJ, 598, 886
\bibitem{} Vigotti M., Carballo R., Benn C.R., De Zotti G., Fanti R., Gonzalez Serrano J.I., Mack K-H, Holt J., 2003, ApJ, 591, 43
\bibitem{} Woo J-H \& Urry C.M., 2002a,  ApJ, 579, 530
\bibitem{} Woo J-H \& Urry C.M., 2002b,  ApJ, 581, L5
\bibitem{} York  D., et al. 2000, AJ, 120, 1579


\end{thebibliography}
\end{document}